# Amateur telescopes discover a kilometre-sized Kuiper belt object from stellar occultation

K. Arimatsu[1,2], K. Tsumura[3], F. Usui[4], Y. Shinnaka[1,5], K. Ichikawa[3,6,7], T. Ootsubo[8], T. Kotani[9,1], T. Wada[8], K. Nagase[8], & J. Watanabe[1]

[1] National Astronomical Observatory of Japan, 2-21-1 Osawa, Mitaka, Tokyo 181-8588, Japan
[2] Astronomical Observatory, Graduate School of Science, Kyoto University, Yoshida-honmachi, Sakyo-ku, Kyoto, 606-8501 Japan
[3] Frontier Research Institute for Interdisciplinary Science, Tohoku University, 6-3 Aramaki Aza-Aoba, Aoba-ku, Sendai, Miyagi, 980-8578, Japan
[4] Center for Planetary Science, Graduate School of Science, Kobe University, 7-1-48, Minatojima-Minamimachi, Chuo-Ku, Kobe 650-0047, Japan
[5] Laboratory of Infrared High-resolution Spectroscopy, Koyama Astronomical Observatory, Kyoto Sangyo University, Motoyama, Kamigamo, Kita-ku, Kyoto 603-8555, Japan
[6] Department of Astronomy, Columbia University, 550 West 120th Street, New York, NY 10027, USA
[7] Department of Physics and Astronomy, University of Texas at San Antonio, One UTSA Circle, San Antonio, TX 78249, USA
[8] Japan Aerospace Exploration Agency (JAXA), Institute of Space and Astronautical Science (ISAS), Department of Space Astronomy and Astrophysics, 3-1-1 Yoshinodai, Chuo-ku, Sagamihara, Kanagawa 252-5210, Japan
[9] Astrobiology Center, National Institutes of Natural Sciences, 2-21-1 Osawa, Mitaka, Tokyo 181-8588, Japan

**Summary paragraph**

**Kuiper belt objects (KBOs) are thought to be the remnant of the early solar system, and their size distribution provides an opportunity to explore the formation and evolution of the outer solar system[1–5]. In particular, the size distribution of kilometre-sized (radius = 1–10 km) KBO represents a signature of initial planetesimal sizes when planets form[5]. These kilometre-sized KBOs are extremely faint, and it is impossible to detect them directly. Instead, monitoring of stellar occultation events is one possible way to discover these small KBOs[6–9]. Hitherto, however, there has been no observational evidence for the occultation events by KBOs with radii of 1–10 km. Here we report the first detection of a single occultation event candidate by a KBO with a radius of ~1.3 km, which is simultaneously provided by two low-cost small telescopes coupled with commercial CMOS cameras. From this detection, we conclude that a surface number density of KBOs with radii exceeding ~1.2 km is ~6 × 10$^5$ deg$^{-2}$. This surface number density favours a theoretical size distribution model with an excess signature at a radius of 1–2 km[5]. If this is a true detection, this implies that planetesimals before their runaway growth phase grow into kilometre-sized objects in the primordial outer solar system and remain as a major population of the present-day Kuiper belt.**

A typical optical brightness of kilometre-sized KBOs (with a heliocentric distance $D = 30 - 50$ astronomical units; 1 au = $1.496 \times 10^8$ km) is expected to be fainter than magnitude ~29 at R band, and they are undetectable even using telescopes with apertures ~10 m. Observations of stellar occultations thus provide a unique opportunity to observe these small objects. Since the duration of a stellar occultation by a kilometre-sized KBO is approximately less than a second[10], observations with time resolutions down to one second or less is essential for its detection. Previous high-cadence photometric studies[11–14] have discovered occultation event candidates by sub-kilometre sized KBOs. However, there has been no detection candidate of occultations by larger, kilometre-sized KBOs, which are expected to be much less frequent events[15]. We have to monitor a large number of stars simultaneously in order to discover these rare occultations. Furthermore, detections of stellar occultation events with ground-based instruments must be robust to terrestrial events such as birds, aircraft, and atmospheric scintillation effects. Therefore, detections with multiple independent telescopes are required for occultation observations from the ground.

To achieve a challenging occultation observation, we launched an observation project using amateur telescopes, Organized Autotelescopes for Serendipitous Event Survey (OASES)[16]. The OASES project uses two identical observation systems (OASES-01 and OASES-02, Supplementary Fig. 1). Each system consists of a 279-mm Celestron, LLC f=2.2 Rowe-Ackermann Schmidt astrograph equipped with a ZWO Co., Ltd. ASI1600 MM-C CMOS camera and a Metabones Speed Booster SPEF-M43-BT4 focal reducer. The effective focal ratio and the angular pixel scale of the observation systems are f/1.58 and 1.96 arcseconds, respectively. The OASES observation systems are capable of monitoring up to ~2000 stars with magnitudes down to $V$ ~13.0 in 2°.3 × 1°.8 field of view simultaneously, providing signal-to-noise ratios comparable to or greater than 3 — 4 with a sampling cadence of 15.4 Hz at an extremely low-cost (~ 16000 USD per a single system). The monitoring observations with the two systems were carried out between 25 June 2016 and 1 August 2017, JST. The two OASES observation systems were installed in different positions on the rooftop of the Miyako open-air school in Miyako Island, Miyakojima-shi, Okinawa Prefecture, Japan (Supplementary Fig. 1a). The separation between the two observation systems was 39 meters (June 2016 ~ June 2017) or 52 meters (July 2017 ~ August 2017). We have selected a monitoring observation field (RA, Dec) = (18:30:00, -22:30:00) very close to the ecliptic (corresponding ecliptic latitude $\beta$ ~ +0°.8) to increase the detectability of the occultation. We thus monitor stars at $\beta = -0°.1 - +1°.6$. During the observation, images of the selected field are obtained with each observation system for a 2 × 2 binned sequential shooting mode of 15.4 frames every second. The exposure time is 65.0 milliseconds for each frame. The details of the OASES observation systems and the monitoring observations are described in ref. 16.

The dataset obtained with the OASES observations in good weather conditions amounts to 60-hour imaging data runs in total, corresponding to approximately 50 terabytes of raw data. Since the relative transverse velocity between the observer and a KBO is dominated by the orbital velocity of Earth, the duration of the occultation by a KBO with a given size depends on a KBO (or occulted star)—Earth—opposition angle[16]. In the present study, we select the data of the selected field observed close to the opposition (with an absolute field—Earth—opposition angle smaller than 30 degrees), corresponding to 42.2-hour data runs. In the selected angle range, the transverse velocity of the KBO on a circular ecliptic orbit relative to the Earth $v_{rel}$ is expected to be $v_{rel} = 20 - 25$ km s$^{-1}$. The occultation shadow size of a kilometre-sized KBO is determined by the radius of the KBO and the diffraction effects, and the minimum diameter of the occultation shadow $W_0$ is approximated with the Fresnel scale $F$, $W_0 = 2\sqrt{3}\ F$[16]. For a KBO, $F$ and $W_0$ correspond to $F$ ~ 1.3 km and $W_0$ ~ 5 km, respectively, at a wavelength of 500 nm. The minimum duration of a KBO occultation $\tau_0$ is thus written by $\tau_0 = W_0/v_{rel}$ and corresponds to be 0.2 – 0.3 seconds for the selected data runs. Therefore we searched for simultaneous flux drops with a duration timescale comparable to $\tau_0$. From 26400 time-sequential images obtained within each data run, we performed aperture photometry for stars with square apertures to produce light curve datasets using a data reduction pipeline developed for the OASES observations[16]. Until now, the obtained light curve datasets include $7.18 \times 10^9$ photometric measurements corresponding to approximately $1.30 \times 10^5$ star hours (the number of observation hours multiplied by the number of observed stars). The detectability of the occultation depends highly on the signal-to-noise ratios of light curves. We thus select light curve datasets with signal-to-noise ratios greater than four. The selected light curve datasets include $3.35 \times 10^9$ photometric measurements corresponding to approximately $6.05 \times 10^4$ star hours (Fig. 1).

Since the OASES two observation systems are capable of high-speed photometry, they are expected to detect an occultation event in two or more flux measurements on their light curves independently. Therefore we searched for consecutive flux drops detected with the two systems simultaneously (see Methods). We discovered one occultation candidate in the dataset of a monitoring field star at an ecliptic latitude of $\beta \sim +0°.2$ (Fig. 2, see also Supplementary Video 1 and Supplementary Fig. 2). The probability of false detection of such an event due to statistical fluctuations is estimated to be $\sim 1 \times 10^{-11}$, and the corresponding expected number of the false detections in our total datasets is 0.03 (see Methods and Supplementary Figs. 3, 4, 5). We should note that the false-positive probability calculations do not rule out an extremely rapid (with a timescale of ~0.1 seconds) and large-scale (larger than ~ 40 meters in physical scale) atmospheric event that affects both observation systems simultaneously. The possible events such as atmospheric scintillation and cloud attenuation occur below the tropopause (corresponding to an altitude of ~ 15000 meters), and the angular scale of these events should be larger than ~ several arcminutes. Therefore similar flux variations due to the atmospheric event would be observed simultaneously for stars at angular distances from the candidate star less than the angular scale. Fig. 3 shows the light curves of the nearby stars obtained at the same time as the event candidate. Simultaneous flux variations were not observed in the light curves of the nearby stars. The present results are consistent with the assumption that the observed event candidate is a true detection. We also note that detections of occultations by the other solar system populations are unlikely. According to the results of a previous ecliptic survey[17] of main-belt asteroids (MBAs), the expected occultation rate of MBAs is about three orders of magnitude less than the present detection rate. The expected occultation rate of the other solar system populations is much less frequent, and a KBO is thus the object most likely to cause the detected occultation.

This is the first detection of an occultation candidate by a kilometre-sized KBO provided by multiple telescopes. Assuming that the spherical occulting object lies on a circular KBO orbit with an inclination of $0°.2$, the best-fit radius yields $1.3^{+0.9}_{-0.1}$ km (Fig. 2, see also Methods). With the reported ecliptic latitude distribution[18] and the effective angular survey area for our occultation observations (Supplementary Fig. 6 and see Methods), our single occultation detection yields a surface number density of KBOs with radii larger than 1.2 km of $5.5^{+12.7}_{-4.6} \times 10^5$ deg$^{-2}$ around the ecliptic (Fig. 4). This surface number density is slightly lower than the upper limit obtained with a previous occultation survey, TAOS[15]. Direct observations of larger-sized KBOs show that there is a possible break of size distribution at a radius of ~45 km[19-21]. Assuming that the size distribution for KBOs with radii smaller than 45 km can be described by a single power law function, i.e., $N(> r) \propto r^{1-q}$, we find $q = 3.9^{+0.3}_{-0.5}$, which is consistent with the previous occultation survey results of sub-kilometre sized KBOs, $q = 3.8 \pm 0.2$ (ref. 12), and $q > 3.5$ (ref. 14), and is marginally comparable to a 95% confidence level upper limit obtained in the TAOS survey, $q < 3.34 — 3.82$ (ref. 15).

At a 95% confidence level, the surface number density with radii larger than 1.2 km ($3 \times 10^4$ — $7 \times 10^6$ deg$^{-2}$, Fig. 4) is greater than that of the modelled scattered disk KBO abundances required to be the source for the observed distribution of the Jupiter family comets[22] ($0.6 — 1.2 \times 10^4$ deg$^{-2}$) and comparable to those of Plutinos[23] and of classical belt objects[24] ($4.5 \times 10^4$ and $7 \times 10^5$ deg$^{-2}$, respectively, assuming a projected sky area for each population to be $10^4$ deg$^2$). Our results thus indicate that the number density of kilometre-sized KBOs is sufficient to supply the nuclei of the Jupiter family comets if their source population is scattered disk. However, we are not able to exclude any of these populations as source populations. On the other hand, our results rule out a simple extrapolation of the size distribution of KBOs observed directly (with sizes down to $r \sim 10$ km)[25] at a 95% confidence level (Fig. 4). This possible discrepancy suggests a second break or an excess of the size distribution at $r = 1 — 10$ km, as predicted by several theoretical models[5,26]. Our present results favour a theoretical size distribution model with an assumption that outer solar system objects are formed from kilometre-sized planetesimals via the runaway growth phase[5] (Fig. 4). According to the model, small planetesimals leave an excess signature in the size distribution of KBOs that corresponds to a typical size of an initial planetesimal population and is not erased after $4.5 \times 10^9$ years of collisional evolution. Our results are consistent with an excess signature of the KBO size distribution at a radius of $1 — 2$ km, which is predicted by the model assuming initial planetesimal radii ranging from 0.4 km to 4 km[5]. However, there is still large statistical uncertainty in the estimated surface number density due to the single event detection in our observations.

To gain a stricter constraint on the size distribution, we plan to carry out further monitoring observations and increase the number of the detected occultations. Our detection promises that on-going and future surveys with multiple high-speed instruments, such as TAOS II[27], CHIMERA[28], and Colibri[29], will provide more detailed and remarkable characteristics of the KBO size distribution via stellar occultations. Our results also demonstrate the potential of optical high cadence observations using amateur telescopes and low-cost CMOS cameras to discover small KBOs. Large occultation surveys, as well as OASES and other small projects by amateur astronomers and citizen scientists will reveal the nature of small outer solar system objects hidden in shadow.

**Acknowledgements**

We thank the personnel of Miyakojima City Museum and of Miyako open-air school (*Miyako Seishonen no Ie*) for providing site infrastructure and access to facilities. We also thank the people of Miyako Island for supporting our observations. We appreciate Dr. Y. Sarugaku for his constructive advices. This research has been partly supported by JSPS grants (JP26247074, 15J10278, 15J10864, 26800112, 16K17796, 18K13584 and 18K13606).

**Author contributions**

KA is a principal investigator of the monitoring campaign described here. KA, KT, FU, TO, TK, TW, KN, and JW developed the observation systems. KA, KT, FU, YS, and KI carried out monitoring observations. KA developed the data reduction pipeline and the occultation detection program with substantial contributions from KT and JW. The figures and movies including the Supplementary Figures and Video were generated by the authors.

**Author Information**

Reprints and permissions information is available at www.nature.com/reprint. The authors declare that they have no competing financial interests. Correspondences and requests for materials should be addressed to KA (ko.arimatsu@nao.ac.jp).

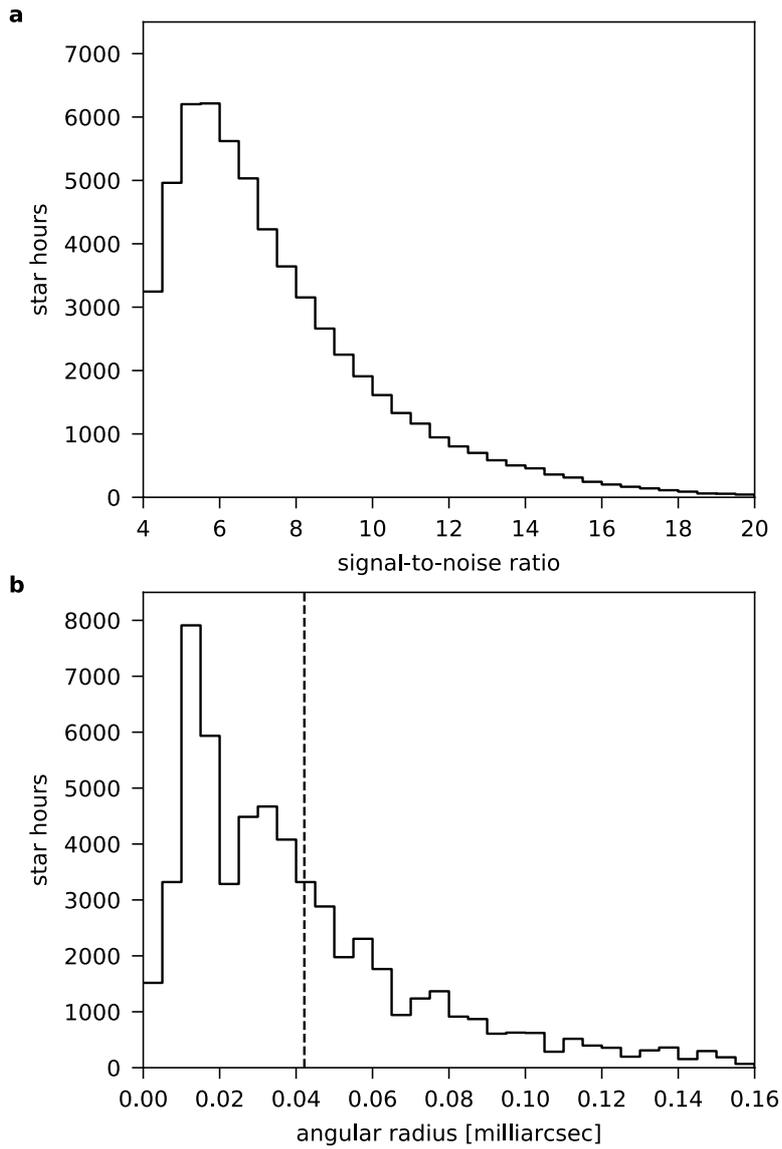

**Figure 1 | Distributions of star hours for the selected light curve datasets used in the present study.** **a**, Distribution of star hours as a function of the signal-to-noise ratio of light curves in a 15.4 Hz bin. **b,** Distribution of star hours as a function of angular radii of the stars. The vertical dashed line indicates the Fresnel scale calculated at 40 AU.

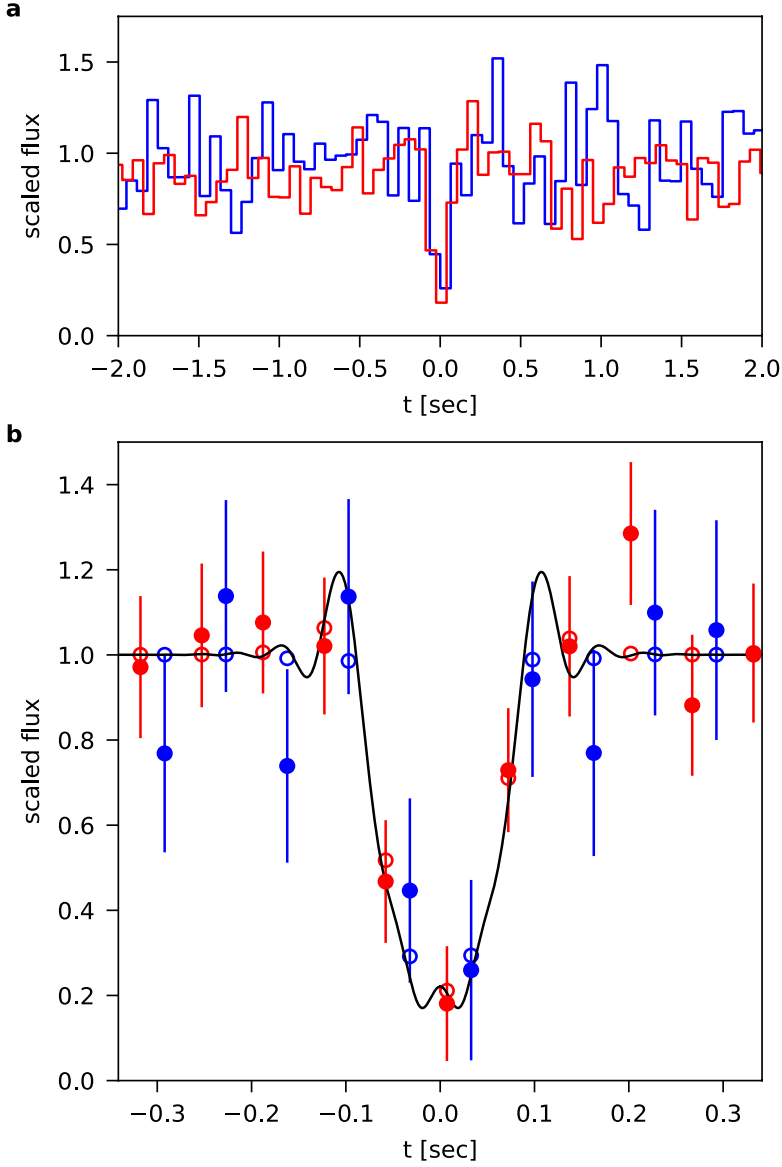

**Figure 2 | Light curves of the occultation event candidate obtained with the two OASES observation systems. a**, the light curves of an occulted star as a function of the time offset $t$ from the central time of the occultation event candidate obtained with OASES-01 (blue line) and OASES-02 (red line), respectively, normalized to average fluxes. The equatorial and ecliptic coordinates of the occulted star are (RA, Dec) = (18:29:02.7, -23:02:34.6) and ($\lambda$, $\beta$) ~ (276°.7, +0°.2), respectively. The Gaia G band magnitude[30] of the star is 12.1. The central time of the occultation candidate is estimated to be June 28th, 2016, 12:56:05.283 UT. The signal-to-noise ratios derived from the light curve of OASES-01 and OASES-02 are 4.9 and 5.4, respectively. For full data set of the light curves, see Supplementary Figure. 2. **b**, Enlargement of the light curves with error bars representing the detector readout noise and the target shot noise overlaid with the best-fit theoretical light curve (black line). The main noise source is the detector readout noise, and typical error bar sizes are ~0.21 and ~0.17 for OASES-01 and OASES-02, respectively. We note that these error sizes are comparable to actual standard deviations of the light curves (0.20 and 0.18 for OASES-01 and OASES-02, respectively). Open blue and red circles correspond to the theoretical light curve integrated over each bin (15.4 Hz interval). Note that the timing of the OASES-01 and -02's exposures are not synchronized. Assuming that the spherical occulting object lies on a circular KBO orbit with an inclination of 0°.2, the best-fit KBO radius, impact parameter, and distance yield $1.3^{+0.8}_{-0.1}$ km, $0.6^{+1.4}_{-0.3}$ km, and $33^{+17}_{-3}$ au, respectively. The best-fit $\chi^2$ value from the fit is 7.0 with 12 degrees of freedom.

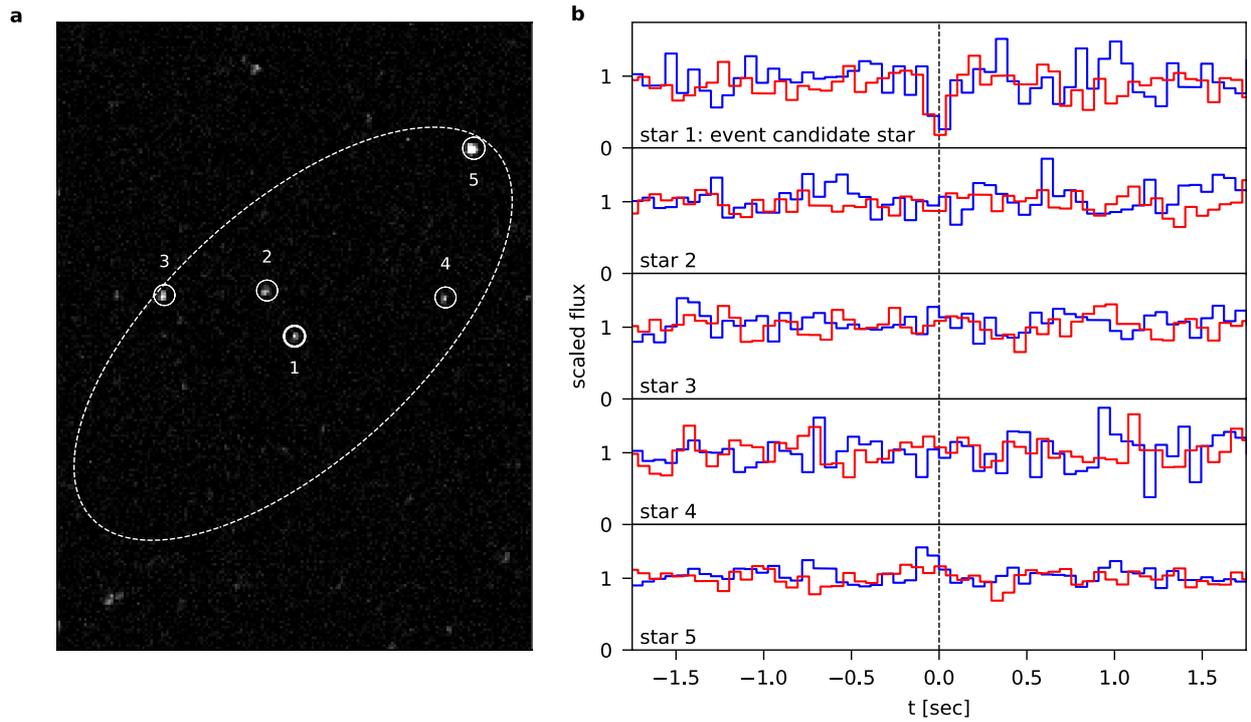

**Figure 3 | Light curves of the nearby stars obtained with the two OASES observation systems. a**, Cutout image of the observation field. The solid circles present positions of the occultation candidate star (star 1) and nearby field stars (stars 2–5). The dashed ellipse represents a region within 40 meters horizontally from the line-of-sight to the candidate star at an altitude of ~ 15000 meters (corresponding to the minimum physical scale and the maximum altitude of an atmospheric event that affects both telescopes simultaneously). **b**, the scaled light curves of the occultation candidate star (star 1) and the nearby field stars (stars 2–5) as a function of the time offset $t$ from the central time of the occultation candidate obtained with OASES-01 (blue line) and OASES-02 (red line). The dashed vertical line represents the central time of the occultation candidate.

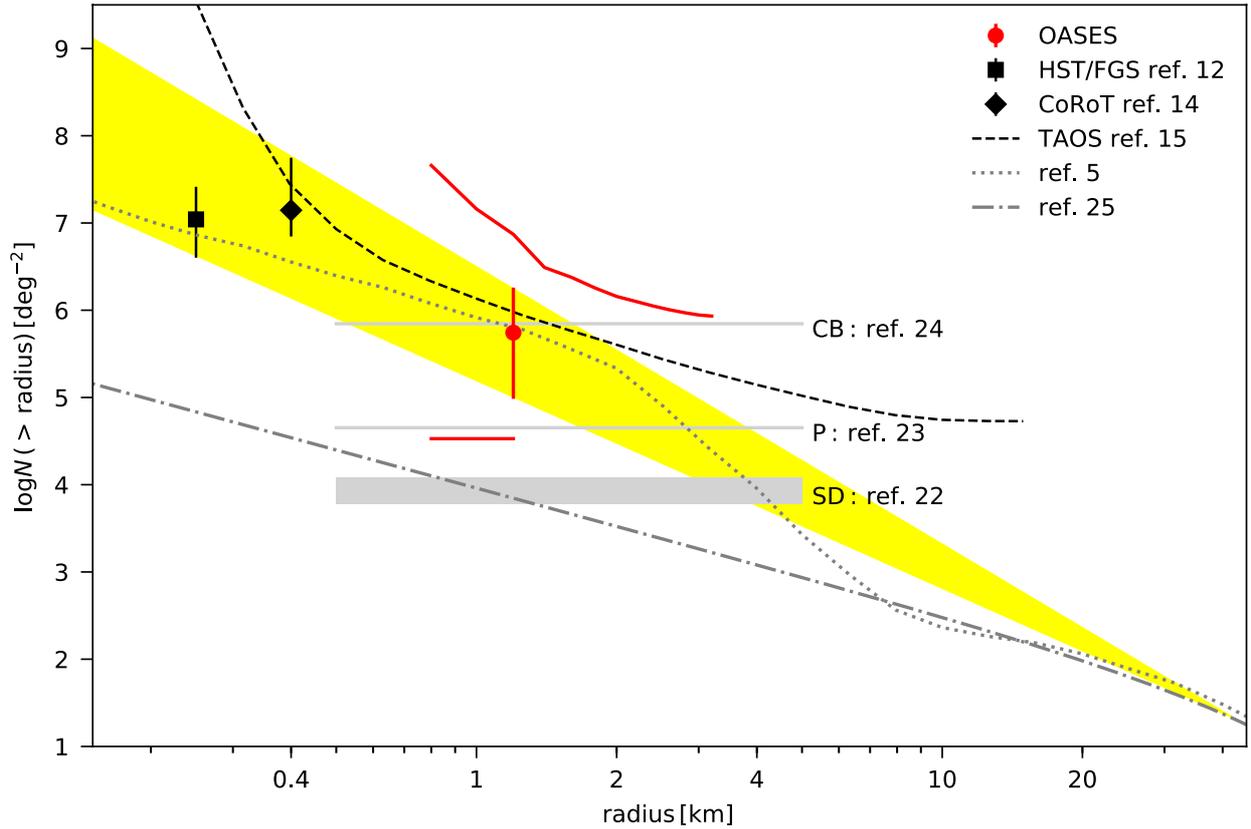

**Figure 4 | Cumulative size distribution of KBOs around the ecliptic as a function of radius.** The red circle is the present results obtained from our OASES detection with a 1$\sigma$ error bar, assuming that kilometre-sized KBOs follow the same inclination distribution as their larger ($r >$ 50 km) populations[18]. The upper and lower red curves correspond to an upper- and lower-limit at a 95% confidence level. The yellow area shows a 1$\sigma$ range for our best estimate of power-law size distribution normalized to $N$ ($r >$ 45 km) obtained with direct searches[25]. We assume that the albedo and the distance of the KBOs discovered by direct searches are 4% and 42 au, respectively, when converting the observed R band magnitude distribution into the size distribution. The black square and the diamond are the results of the previous optical occultation surveys with the Hubble space telescope's Fine Guidance Sensors[12] and the *CoRoT* space telescope[14], respectively. In addition, a 95% upper limit from the TAOS occultation survey[15] is shown as the dashed line. The grey lines and area labelled CB, P, and SD are theoretical estimates based upon models of the classical belt[24], Plutinos[23] and scattered disk KBOs[22], respectively, as a source of the Jupiter family comets. The surface densities of the classical belt and Plutinos are derived by assuming a projected sky area to be $10^4$ deg$^2$ for each population. The dotted line is a theoretical collisional evolution model assuming initial planetesimal radii ranging from 0.4 km to 4 km[5]. The dotted-dashed line represents the extrapolation of the fitted size distribution of KBOs obtained by direct searches ($r >$ 10 km) [25].

**Methods**
**Detection algorithm**

In several light curve datasets, low-frequency (with timescales larger than several seconds) flux variations are seen. These variations are most likely due to changes in atmospheric transparency and could result in degrading detection performance of occultations. To remove these variations, a numerical high pass filter with a cut-off time scale of seven times of the minimum occultation duration timescale $\tau_0$ has been applied to light curve datasets. We should note that the choice of the cut-off timescale is not critical for the present results, as long as it is sufficiently larger than $\tau_0$ and smaller than that of the typical low-frequency flux variations ($\sim 20\ \tau_0$).

In light curve datasets for each star, typically consisting of $\sim 26000 \times 2$ flux bins, we search for occultation event candidates using "rank statistics method"[31]. In this method, one calculates the descending order of rank of the flux at each time bin in a light curve dataset and compares the products of the ranks obtained with individual systems $i$ at the same time bin $j$, $r_{ij}$. If a flux drop due to an occultation occurs in the light curves simultaneously, one can detect the occultation as an extraordinary large value of the sum of the logarithmic rank product at a time bin $j$, $\eta(j)$ (see ref. 31) defined as follows,

$$\eta(j) = -\ln\left(\frac{\prod_{i=1}^{T} r_{ij}}{N^T}\right), \qquad (1)$$

where $N$ is the number of time bins in a light curve dataset for each star and each observation system and $T$ is the number of observation systems ($T = 2$ for the present study). This method was originally developed for the detection of simultaneous flux drops and the calculation of their statistical significances in the light curves obtained with the multiple independent instruments. However, the number of instruments currently used for the OASES observations ($T = 2$) is insufficient to detect any event with high statistical significance using this method[31]. Therefore we only use this method as a detection method of simultaneous flux drops. For the rank statistics method, we perform a smoothing process of the light curve dataset using a moving average filter with a window whose width corresponds to $\tau_0$. This smoothing process makes it possible to detect consecutive flux drops efficiently. Due to extremely rapid fluctuations of atmospheric conditions, several data runs could suffer correlations between two system's light curves which would possibly give rise to a larger number of false detections in the rank statistics method[31]. We thus perform correlation tests on the smoothed light curve datasets with their signal-to-noise ratios greater than 10 for each data run, as proposed by ref. 31. We currently use the light curve datasets obtained in the data runs that pass the tests. In these datasets, we searched for occultation candidates with $\eta$ greater than 14. Furthermore, to find candidates with sufficient statistical significances, we selected light curves of event candidates with $\chi^2$ values calculated for the fit of a flat light curve (corresponding to a light curve of no event) greater than 10 for each system and 25 for the sum of the two systems' values. We found 2431 candidates in total.

In the candidates, there is a large number of false detections due to photometry of images showing time variations of their point spread functions. These variations are caused by instantaneous tracking imperfections and result in false flux estimates. For the occultation candidates detected in the rank statistics method, we re-performed aperture photometry using larger circular apertures to produce light curves without the effect of the point source variations. With the reproduced light curves, we select candidates according to the following criterion;

    (1) flux drops with their p-values obtained from $\chi^2$ test results for the fit of a flat light curve (corresponding to p-values of the individual groups of flux measurements) smaller than $3.3 \times 10^{-4}$ for each system.

This statistical significance corresponds to the expected significance for the occultation event by a 0.8 km KBO in a light curve with its signal-to-noise ratio of $\sim 5$. Ten event candidates pass the test.

For the candidates, we check the imaging data of the event candidate stars whether the flux drops are due to severe instantaneous tracking errors possibly caused by strong winds or confusions with nearby sources and artifacts. After that, assuming that an identical occultation event with a duration longer than a single time bin is detected with the two systems simultaneously, we check the event candidates using the light curves normalized to average fluxes (Fig. 2a) and a time window centred at the time of the maximum flux drop in the OASES-02 light curve with a width of $\tau_0$ according to the following additional criteria;

> (2) For each system's light curve, two or more flux measurements in the time window show flux drops with statistical significances greater than $2\sigma$.
> 
> (3) Two light curves obtained with the two systems are well synchronized with the p-values obtained from a $\chi^2$ test for comparison between the flux measurements of OASES-01 and those of OASES-02 in the same window less than 0.05.

We found one occultation event candidate that satisfies these criteria (Fig. 2 and Supplementary Video 1., see also Supplementary Fig. 2).

**False-positive estimate**

We calculate $\chi^2$ values for the fit of a flat light curve to flux measurements of the event candidate in the time window used for the detection algorithm (corresponding to two and three measurements for OASES-01 and OASES-02, respectively). In this calculation, we use a light curve normalized to the average value of flux measurements outside the time window and within $7\tau_0$ (corresponding to ~0.6 sec, the same duration of the moving average filter used in the detection algorism) before and after the window central time. The obtained $\chi^2$ values are 19.7 with two degrees of freedom for OASES-01 and 25.2 with three degrees of freedom for OASES-02. The corresponding p-values (p-values of the individual groups of flux measurements $p_1$ and $p_2$ for OASES-01 and -02, respectively) are $p_1 = 5.20 \times 10^{-5}$ and $p_2 = 1.38 \times 10^{-5}$. The dominant noise source of the two light curves is expected to be random detector readout noise. Supplementary Figure 3a and b show the p-value distributions for the light curve obtained with the OASES-01 and -02, respectively. No clear trend is apparent in each distribution. Therefore we assume that the light curve datasets of the occultation event candidate are independent and identically distributed. We derive the combined p-value of the event candidate using Fisher's method. Assuming that the two p-values obtained at the same time bin, $p_1$ and $p_2$, derived from independent and identically distributed datasets, the distribution of -2 times of logarithm sums, $-2(\ln(p_1) + \ln(p_2))$, follows the $\chi^2$ distribution with four (corresponding to two times of the number of the datasets) degrees of freedom. Using this Fisher's method, the combined p-value of the event candidate is derived to be $1.58 \times 10^{-8}$. Supplementary Figure 3c shows the combined p-value distribution for the observed two system's light curves. No clear trend is apparent in the distribution. This distribution is consistent with the assumption that the obtained p-value for the event is reasonable. For random fluctuations with their p-values equal to or smaller than that for the event candidate, the probability of producing flux drops that satisfy the detection criteria (1), (2) and (3) is estimated to be $\sim 6 \times 10^{-4}$. Among these three criteria, criterion (2) is the strictest constraint. Randomly generated fluctuations including flux increase events, flux drops with short durations, and those with long durations but with a significant flux drop measured only in a time bin are rejected by this criterion. Even if flux drop events that pass criterion (2) are obtained by the two systems with sufficiently small combined p-values, those with an insufficient significance for one of the systems and those with different timings and depths are rejected by the other criteria, and the probability thus becomes small. The probability of a false detection due to a chance coincidence of the random fluctuations is $\sim 9.0 \times 10^{-12}$, and the expected number of the false detections in our total datasets is approximated to be 0.03.

In case that the combined p-value distribution has an excess at the low end due to an irregular condition, the combination of the obtained p-values ($p_1$ and $p_2$) can be underestimated. In order to address the concern, we carried out a test using the light curve datasets. Supplementary Figure 4 shows the histogram of the combined p-values of observed simultaneous flux drops overlaid with the distribution expected from Fisher's method. To investigate the characteristics of the light curves under similar atmospheric conditions, we use the light curve datasets obtained within 20 minutes before and after the occultation event candidate. Since the obtained distribution has no excess feature at lower p-values, we found no evidence of underestimation of the combined p-value of the event candidate.

In addition, we estimate the possibility of the false detection at the extreme lower p-value end to address the concern that the event candidate was a false positive detection due to a temporary condition. We select light curve datasets of stars obtained within two hours before and after the occultation event with their signal-to-noise ratios of 4 — 6 (comparable to those of the event candidate, ~5). The number of the corresponding flux measurements obtained with each system is $2.2 \times 10^7$. Assuming that the datasets are independent, we permuted the flux measurements obtained with OASES-01 to those obtained with OASES-02 and simulated $4.8 \times 10^{14}$ flux measurement pairs to determine the probability of the false positive detections due to random noise, as shown in Supplementary Fig. 5. In these simulated measurements, we found $4.6 \times 10^3$ flux drop events that satisfy the detection criteria with their combined p-values less than that of the observed occultation event candidate ($1.58 \times 10^{-8}$). The probability of simultaneous flux drops due to random fluctuations is thus approximated to be $(4.6 \times 10^3) / (4.8 \times 10^{14}) \sim 9.5 \times 10^{-12}$, which is consistent with that derived by the previous false positive estimate ($\sim 9.0 \times 10^{-12}$).
From these test results mentioned above, we conclude that the probability of our occultation event candidate arising from random fluctuations within our dataset is approximately three percent.

**Light curve fit**

In addition to the size, the impact parameter, and the distance of the KBO, we also consider the angular diameter and the spectrum of the occulted star in the calculation of the theoretical light curve (Fig. 2b). To estimate the angular diameter and the spectrum of the occulted star (Gaia DR2 source id: 4089842750319814784), we carried out spectral model fit with the following flux catalogue data; Gaia DR2[30] BP ($m_{BP}$ = 12.312 ± 0.006) and RP ($m_{RP}$ = 11.757 ± 0.004) bands and 2MASS[32] J ($m_J$ = 11.304 ± 0.023), H ($m_H$ = 11.183 ± 0.021), and Ks ($m_K$ = 11.165 ± 0.023) bands. The comparison of these catalogued values with a stellar spectrum model[33] indicates that the catalogued fluxes are explained by spectral models with an effective temperature between 7000 and 7250 K and log $g$ between 4.5 and 5.0. We thus select a spectral model with $T_{eff}$ = 7000 K, log $g$ = 5.0, and [$M/H$] = 0.0 (solar metallicity) as the template stellar spectral energy distribution. The best-fit parameters yield a stellar angular diameter of 0.0220 ± 0.0003 milliarcseconds and an extinction value of $E(B - V)$ = 0.070 ± 0.004. $T_{eff}$ for the selected model and the best-fit stellar angler diameter and $E(B - V)$ values are consistent with those from the Gaia DR2 catalogue[30]. With the best-fit parameters, the selected stellar spectral model, and the system efficiency of the OASES camera unit[16], the theoretical light curve for the fit is calculated with a model based on the occultation simulation described in ref. 10. The selection of the stellar spectrum model does not affect the theoretical light curve (smaller than one percent of the intensity of the theoretical light curve integrated over each time bin). The best-fit results are shown in Fig. 2b.

Since the GPS time synchronization of the current OASES systems did not work perfectly, we carried out an additional timing calibration using meteors that appeared in the observation data (see ref. 16). After the calibration, the timing between the two systems is synchronized with the 1 σ accuracy of eight milliseconds, corresponding to ~ 1/10 of the single-frame exposure time. This timing uncertainty is not critical for the present fit results. The best-fit parameters (radius $r$, impact parameter $b$, and distance $D$) for individual light curves are consistent with those derived from the two light curves, yielding $r = 1.2^{+1.8}_{-0.3}$ km, $b = 0.7^{+2.4}_{-0.7}$ km, and $D = 31^{+19}_{-1}$ au for OASES-01 and $r = 1.4^{+1.4}_{-0.2}$ km, $b = 0.7^{+2.0}_{-0.4}$ km, and $D = 31^{+19}_{-1}$ au for OASES-02, respectively. The most likely distance estimated using the light curve of each observation system is slightly smaller than that of the two systems ($D = 33$ au). Since the time difference between the two individual detections is not considered, the duration of the occultation becomes shorter than that of the two light curves, which makes the estimated $D$ slightly smaller.

**Kuiper belt surface number density**

With the size distribution of KBOs with a radius $r$, $N(r)$, the expected number of detected occultations $N_{exp}$ is expressed by

$$N_{exp} = \int_{r_{min}}^{r_{max}} \Omega(r) \frac{dN(r)}{dr} dr, \quad (2)$$

where $\Omega(r)$ is an effective angular survey area of the OASES observations for KBOs with a radius $r$. $\Omega(r)$ can be written as

$$\Omega(r) = \varepsilon(r)|f(\beta)|\frac{W(r)}{D}\sum_{l}\frac{v_{rel}}{D}t_l, \quad (3)$$

where $\varepsilon(r)$ is the detection efficiency of an occultation by a KBO with a radius $r$, $|f(\beta)|$ is the averaged value of the ecliptic latitude distribution of KBOs, $f(\beta)$, over the selected observation field, $W(r)$ is the diameter of the KBO occultation shadow[16], $v_{rel}$, and $D$ are the relative velocity and distance to the KBO, respectively, and $t_l$ is the size of each light curve bin $l$ in time. In this study, we assume that $f(\beta)$ is independent of $r$ and follows that of larger ($r > 50$ km) KBOs[18]. We also assume that $D$ is 40 au as a typical KBO distance. We estimate $\varepsilon(r)$ by recovering theoretical light curves of occultations by a KBO with a radius ranging from $0.8 < r < 3.2$ km implanted in randomly selected actual light curve data. The impact parameter of the theoretical light curves ranges from zero to one time the occultation shadow size $W(r)$. The stellar spectral energy distribution and the angular stellar size used to produce each theoretical light curve are determined by the spectral model fit to the flux catalogue data (see Methods) of the randomly selected star. For example, the estimated $\varepsilon(r)$ for $r = 0.8$, 1.2, and 2.0 km are 0.03, 0.2, and 0.6, respectively. Supplementary Fig. 6 shows the estimated $\Omega(r)$ as a function of $r$. According to the obtained $\Omega(r)$, an occultation by a KBO with $r > 1.2$ km can be efficiently detected with the OASES datasets.

Assuming that the KBO size distribution follows a power law function, $N(r) = n_0 \times r^{1-q}$, $N_{exp}$ can be written as

$$N_{exp} = n_0(q-1)\int_{r_{min}}^{r_{max}} \Omega(r) r^{-q} dr, \quad (4)$$

From equation (4), we calculate a cumulative KBO surface number density around the ecliptic (-5° < $\beta$ < 5°). In the present study, we assume $q = 4.0$ and derive $N(r > 1.2$ km) to be $5.5 \times 10^5$ deg$^{-2}$. However, we should note that the derived cumulative surface number density at $r = 1.2$ km is not significantly dependent on $q$. For example, $N(r > 1.2$ km) ranges from $5.5 \times 10^5$ to $5.9 \times 10^5$ for $2.5 < q < 4.5$.

**Data availability**

The imaging data frames of the detected occultation are available from https://www.dropbox.com/s/nax9gosyrxbye0o/imgfile.zip.

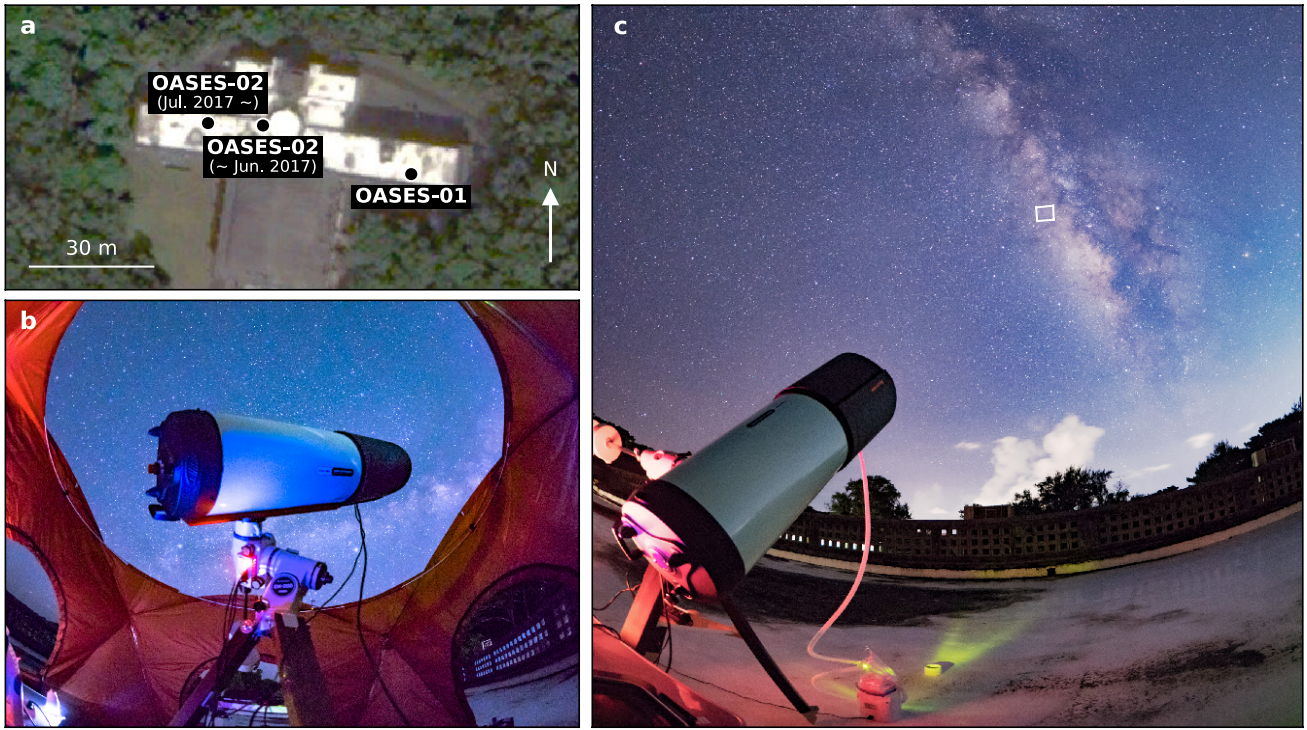

**Supplementary Figure 1 | Observation site and two OASES observation systems. a**, satellite image of the observation site (Miyako open-air school, latitude: 24° 48' 17"N, longitude: 125° 18' 55"E, altitude: 33 m) overlaid with the layout of the two OASES observation systems (OASES-01 and OASES-02). Base imagery from Google Earth Pro v7.3.1.4507 (map data: Google LLC, image data: ©2018 DigitalGlobe) and modified in Matplotlib v.2.0.2. **b**, OASES-01 observation system. OASES-01 was installed in a portable dome-shaped enclosure. **c**, OASES-02 observation system and the location of the monitoring observation field selected for our study (the sky region surrounded by the white lines, (RA, Dec) = (18:30:00, -22:30:00)). OASES-02 was not installed in the enclosure due to lack of financial resources.

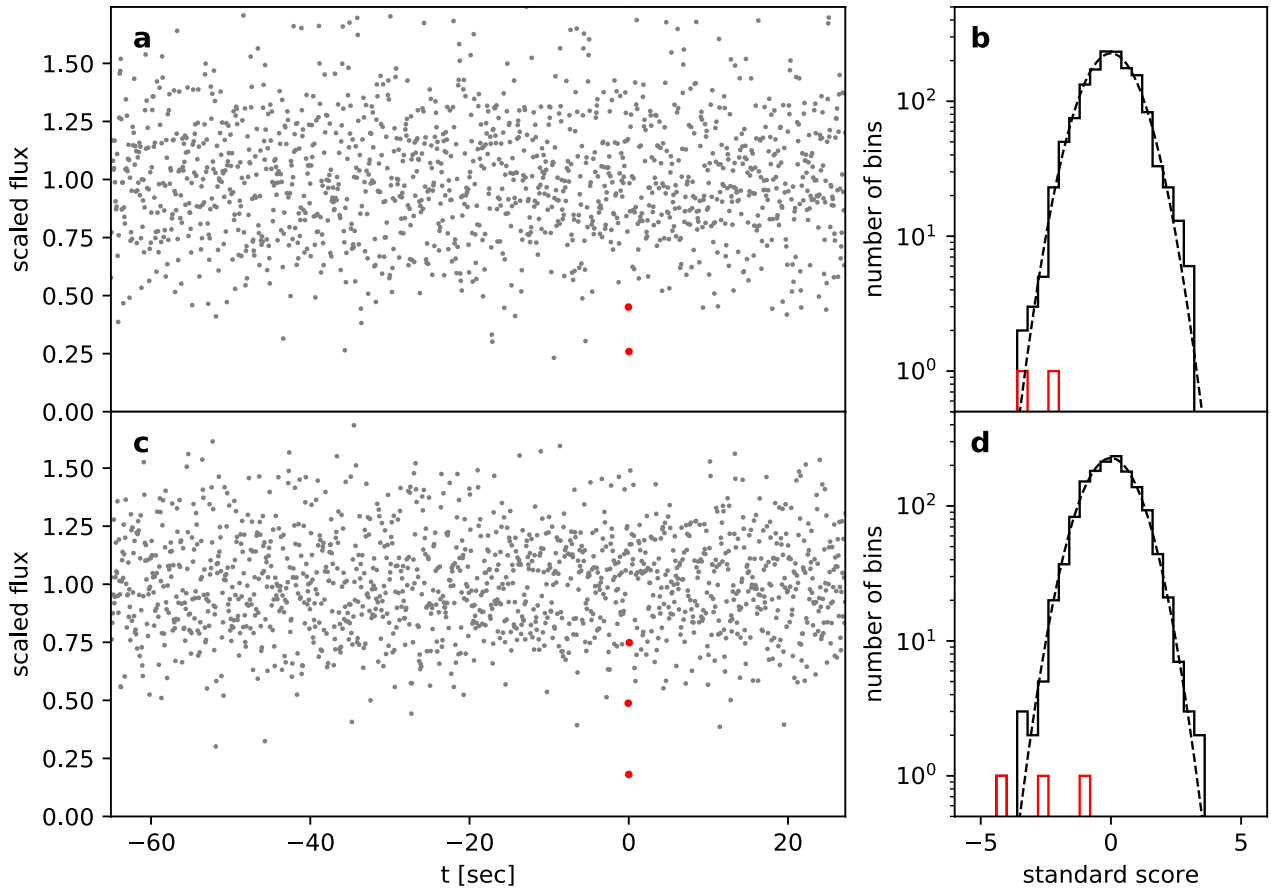

**Supplementary Figure 2 | Full set of sequential photometry of an occulted star.** The scaled flux of an occulted star as a function of the time offset t from the central time of the occultation candidate obtained with OASES-01 (**a**) and OASES-02 (**c**). The red points in each panel correspond to the flux measurements of the event candidate in the time window used for the detection algorithm (see Methods). **b** and **d**, the standard score distribution of flux measurements shown in **a** and **c**, respectively. The standard score for each flux measurement is calculated by subtracting the mean value from the flux and dividing by the standard deviation. The solid black and red histograms are the distribution of all flux bins and that of the event candidate bins, respectively. The dashed line corresponds to a Gaussian distribution.

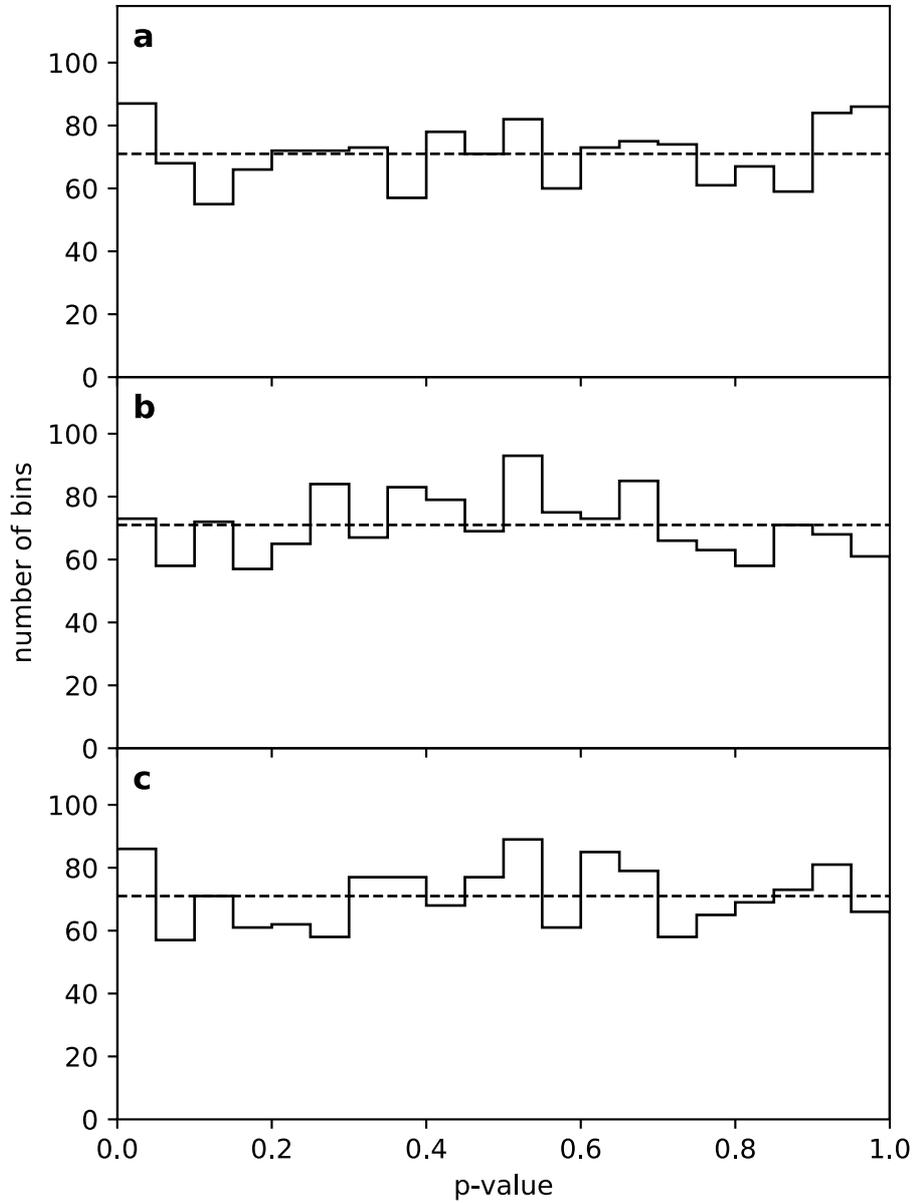

**Supplementary Figure 3 | p-value distribution for the observed light curves.** The solid line histograms show the p-value distributions of the individual light curves of the occultation event candidate (see Supplementary Fig. 2) obtained with the OASES-01 (**a**) and -02 (**b**) systems, and that of the combined p-values using Fisher's method (**c**), respectively. The dashed line in each panel represents the flat distribution.

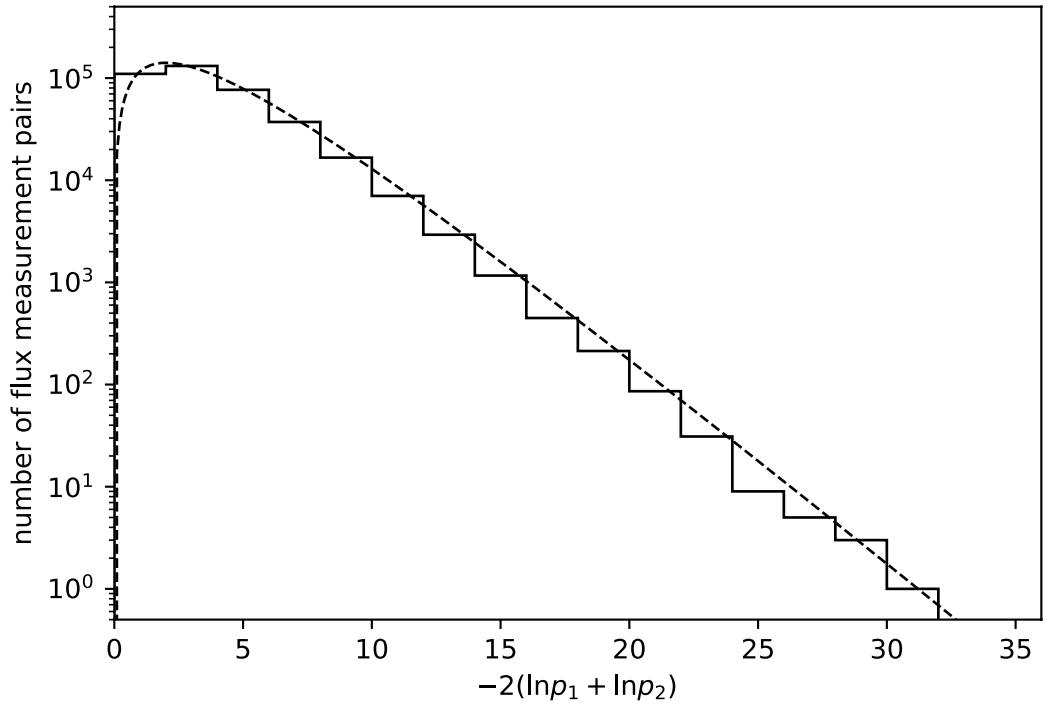

**Supplementary Figure 4 | Distribution of combined $\chi^2$ values of the flux drop bins in the light curves.** The solid line histogram shows the distribution of the Fisher's logarithmic combined probability of the flux drop events observed in the two system's light curves (see Methods). The dashed line represents the distributions expected from the Fisher's method (corresponding to a $\chi^2$ distribution with four degrees of freedom). We use $3.8 \times 10^5$ flux drop events of stars observed within 20 minutes before and after the occultation event. The signal-to-noise ratios of the observed stars are 4 – 6 (comparable to those of the event candidate, ~ 5). We removed the occultation event itself from the dataset.

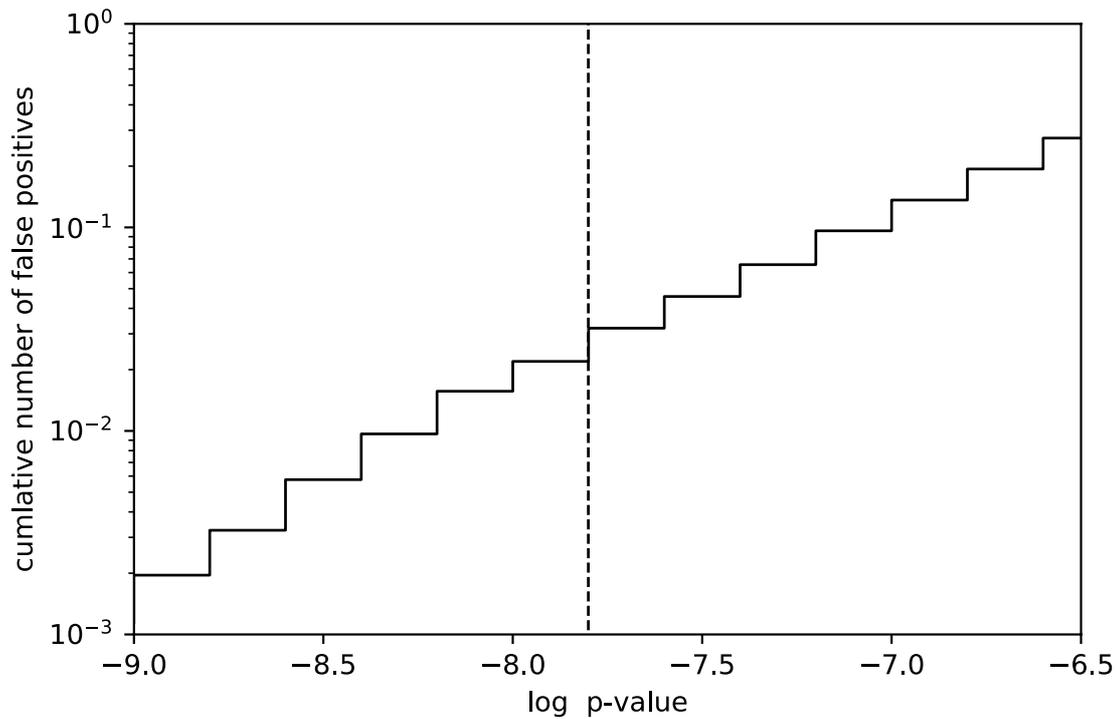

**Supplementary Figure 5 | Cumulative number of false-positives as a function of *p-value.*** The probability of false-positive detections due to random flux fluctuations are obtained using light curve datasets of stars obtained within 2 hours before and after the occultation event with their signal-to-noise ratios of 4 — 6 (see Methods). In the entire analysis, we simulated $4.8 \times 10^{14}$ flux measurements using light curve datasets. These false positive detections pass the detection criteria described in Methods. We removed the occultation event itself from the datasets. The vertical dashed line indicates the corresponding p-value of the observed event candidate ($1.58 \times 10^{-8}$). In the entire simulated measurements, we found $4.6 \times 10^{3}$ events with p-values less than $1.58 \times 10^{-8}$. The probability of simultaneous flux drops due to random fluctuations is thus approximated to be ~ $9.5 \times 10^{-12}$ and the expected number of false detections in the datasets is deduced to be ~0.032.

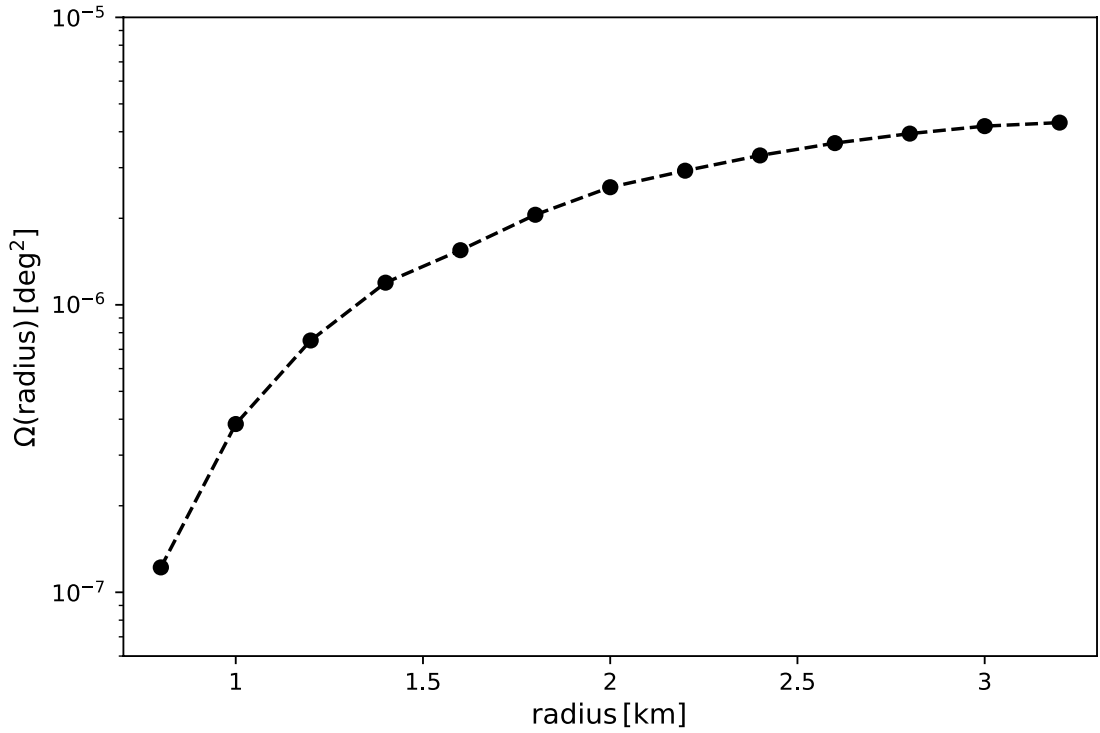

**Supplementary Figure 6 | The estimated effective angular survey area $\Omega(r)$ of the OASES two-year occultation observations for KBOs as a function of KBO radius.** We derive $\Omega(r)$ with the detection efficiency for an occultation by a KBO with each radius estimated by recovering theoretical light curves of occultations by a KBO located at a distance of 40 au with different impact parameters and sub-frame offsets (see details in Methods). The theoretical light curves are implanted in randomly selected actual light curve data and are recovered using the detection algorithm. The stellar spectral energy distribution and the angular stellar size used to produce each theoretical light curve is determined by the spectral model fit to the flux catalogue data (see Methods) for the star of the selected light curve.

**Supplementary Video 1 | Movie data of the occultation event candidate obtained with the two OASES observation systems.** Cutout movies obtained with **a**, the OASES-02 and **b**, the OASES-01 observation system, respectively. The reproduction speed corresponds to 0.05 times of the recording speed. The recorded time (UT) for each frame after the timing correction (see ref. 16 and Methods) is also shown at the bottom of each panel. The output light curves (blue and red points with error bars for OASES-01 and -02, respectively) and the best-fit theoretical light curve (black line, see also Fig. 1b) at the corresponding time are also shown in **c**. Point sources in the OASES-01 images (**b**) extend vertically due to imperfect adjustment of the optical instrument. The full-width half maximum of stars in the images is typically 15 and nine arcseconds for OASES-01 and OASES-02, respectively. The white circle in each panel presents a circular aperture with a diameter of 31 arcseconds used for photometry.